\documentclass[onecolumn,showpacs,showkeys,preprintnumbers,amsmath,amssymb]{revtex4}

\usepackage{graphicx}
\usepackage{dcolumn}
\usepackage{bm}

\begin{document}


\title{ Exact classical Doppler effect derived from the photon emission process}


\author{Chyi-Lung Lin } 
\email{cllin@scu.edu.tw}
\author{Jun-Jia Pan}
\author{Shang-Lin Hsieh}
\author{Chun-Ming Tsai}
\affiliation{Department of Physics,\\ Soochow University, \\Taipei,Taiwan, R.O.C.}






\begin{abstract}
{ 
The concept of photon is not necessary only applied to the  relativistic Doppler theory. It may also work well for classical theory.
As conservation of momentum and energy are physical laws, 
if applying these laws gives the exact relativistic Doppler effect,
it should also give the exact classical Doppler effect. 
So far the classical Doppler effect is only obtained by using some approximation, as derived by Fermi in 1932. We show that the exact classical Doppler effect can be derived from the photon emission process in the exact treatment and reveal that these results are the same as those derived from the wave theory of light.
}
\end{abstract}





\maketitle


\section{\label{sec:INT}  Introduction}

Doppler effect discusses the difference between the emitted and received frequencies of radiation in cases that there is a relative motion of the source and the detector.
The derivation is shown in textbooks with the wave theory of light. 
When the photon theory of light was introduced, it was then necessary to examine the Doppler effect from photon theory. We may study Doppler effect from an excited atom which decays and emits a photon.
A photon is a corpuscular theory of light, treated as a particle with momentum $\vec{P}$ and energy $E$. By wave-particle duality, the momentum and energy of a photon are related to its associated wave properties by $E = \hbar \omega = h \nu $, and $\vec{P}= \hbar \vec{k}$, where $\vec{k}$ is the wave vector and $\nu$ the frequency ($\omega$ the angular frequency), and $c=\omega/ k$, the speed of the wave and the photon.

The investigation of the photon Doppler effect began with the relativistic theory. 
Schr\"odinger in 1922 considered the emission of a light quantum from a moving excited atom and applied the laws of conservation of momentum and energy. He seemed to be the first towards this direction of investigation \cite{b.r1}.
References \cite{b.r2, b.r3} give a more detailed discussions on this work of Schr\"odinger.
Then there followed some formal derivations of the relativistic Doppler effect from the photon emission process. \cite{b.r4, b.r5, b.r6, b.r7, b.r8}.  In the appendix, we also offer a derivation of the relativistic Doppler effect for pedagogical interest.

For the photon emission, we consider an excited atom  $A^*$ which decays to a lower energy state, $A$, accompanied with a photon $\gamma$. The conservation of momentum and energy in the photon emission process is expressed as:  
\begin{eqnarray}
&& 
A^* \rightarrow A + \gamma.
\nonumber
\\
\label{Eq.1-1}
&& 
\vec{P}_{A^*} = \vec{P}_A
+\vec{P}_\gamma.
\\
\label{Eq.1-2}
&&
E_{A^*} = E_A + E_\gamma. 
\end{eqnarray}

\noindent
Previously, the relativistic Doppler effect was considered first. The momentum $\vec{P}$ and energy $E$ of the atom are expressed in a relativistic way. Together with the success of Compton scattering explained by the photon theory and the relativity, hence,  we generally have the impression that photon is a concept only suited for relativistic theory.
 
However, the concept of photon is not necessary only applied to the relativistic theory.
The concept of photon also works well for classical theory.
As conservation of momentum and energy are physical laws, if applying these laws gives the exact Doppler effect in the relativistic theory, it should also give the exact Doppler effect for the classical theory. 

There were already some discussions on the classical Doppler effect in terms of photon theory. 
Fermi in 1932 discussed the emission of a light quantum from a moving atom. The  classical Doppler effect was deduced from the conservation of momentum and energy in the emission process; however, he used some approximation.  
\cite{b.r9}. We discuss this at the end of Sec. II.
There were also similar papers obtaining the classical Doppler effect by approximation \cite{b.r10, b.r11}. 

So far as we know, there is no classical Doppler effect derived from the photon theory in the exact treatment. 
We note that Fermi got the exact classical Doppler effect by using approximation, we may then wonder if the same exact result can be obtained without any approximation.
Thus, it would be interesting, and also  pedagogically interesting, to investigate whether the exact classical results can be derived from the photon emission process in the exact treatment.

In fact, relativistic Doppler effect is comparatively simpler, as there is no need of a medium.
For classical wave theory, we need to introduce a medium for the propagation of waves. The medium serves as an absolute reference frame. We define such a reference frame in which the transmitting medium is at rest as $K$.
This $K$-frame is therefore the absolute reference frame.
The propagation speed of waves and also photons is defined as $c$ in the frame $K$. 
In classical wave theory, we also need to distinguish whether it is the source that is moving or the detector that is moving or both are moving. 

Below, we list the results of Doppler effect derived from the wave theory of electromagnetic waves described in the book by Jackson \cite{b.r12}.
We have the following results. 
For two reference frames $S$ and $S'$ with a relative velocity $\vec{v}$. The relation of the  position vector of these two frames is $\vec{x}'=\vec{x} - \vec{v} t$. Then the unit vector, $\hat{n}$, describing the propagation direction of light waves,  the angular frequency $\omega$, and the propagation speed $c$ in these two frames are related by

\begin{eqnarray}
&& 
\hat{n}' =\hat{n},
\label{Eq.1-3}
\\
&& 
\omega' = \omega (1-\frac{\hat{n} \cdot \vec{v}}{c}),
\label{Eq.1-4}
\\
&&
c'= c- \hat{n} \cdot \vec{v}.
\label{Eq.1-5}
\end{eqnarray}
As the wave vector $\vec{k} = (\omega/c ) \hat{n}$, Eqs.~(\ref{Eq.1-3})-(\ref{Eq.1-5}) also implies that 
\begin{equation}
\label{Eq.1-6}
\vec{k}' = \vec{k}.  
\end{equation} 
These formulas are to be compared with those derived from the photon theory.

We suppose that the photon in the classical theory, called the classical photon, is still described as a particle associated with the wave-particle duality. Thus, the momentum and the energy are described as: 
$E = \hbar \omega $, and $\vec{P}= \hbar \vec{k}$, and $c=\omega/ k$ is the speed of the wave and also of the photon. 
Then, the definition of a photon is parallel in the relativistic and in classical theory. The only difference is that the speed $c$ is not a constant in the classical theory. 

For the notations used, we denote by  $q _{(s)}$ as quantities that are measured in the source frame, and  $q _{(d)}$ in the detector frame. Thus, $\vec{P}_{A(d)}$
represents the momentum of the atom $A$ measured in the detector frame, $E_{A^*(s)}$ the energy of the atom $A^*$ in the source frame, $\nu_{(s)}$  the frequency measured in the source frame and $\nu_{(d)} $ the detector frame. We are to obtain Doppler effect connecting $\nu_{(s)}$ and $\nu_{(d)}$.

We first discuss the case of a moving source in Sec. II, and then discuss a moving detector in Sec. III.
In Sec. IV, we discuss the relations of Doppler shift between two arbitrary inertial frames, and we compare these results with those derived from the wave theory. In Sec. V, we have a conclusion.

\section{\label{sec:II} The source of light is moving and the detector is at rest}

We first consider the case that the detector is stationary in the absolute frame $K$, while the source of light, $A^*$, is moving with a velocity $\vec{v}$ relative to the frame $K$. 

In the detector frame, $\vec{v}_{A^*(d)} = \vec{v} $.
And we let $\epsilon$ represent the energy difference of the excited and lower energy level, and let $m$ be the mass of the atom. 
Eqs.~(\ref{Eq.1-1})-(\ref{Eq.1-2}) can be expressed as follows: 
\begin{eqnarray}
&&
A^* \rightarrow A + \gamma  
\nonumber
\\
\label{Eq.2-1}
&&
 m \vec{v} = m \vec{v}_{A(d)}
 +\vec{P}_{\gamma (d)},
\\ 
\label{Eq.2-2}
&&
\epsilon+ \frac{m}{2} {\vec{v}}^2=  \frac{m}{2}
 {\vec{v}_{A(d)}}^2 
 +E_{\gamma(d)},
\end{eqnarray}
where  
$E_{\gamma(d)} = h \nu_{(d)}$, and 
$\vec{P}_{\gamma(d)} =( h \nu_{(d)}/c) \hat{n}$.
Note that the speed of the light quanta is $c$ in the detector  frame, as the detector frame is the absolute frame.

In the source frame, $A^*$ is at rest. Eqs.~(\ref{Eq.1-1})-(\ref{Eq.1-2}) in the source frame are expressed as

\begin{eqnarray}
&&
A^* \rightarrow A + \gamma  
\nonumber
\\
\label{Eq.2-3}
&&
 0 = m \vec{v}_{A(s)}
 +\vec{P}_{\gamma (s)},
\\ 
\label{Eq.2-4}
&&
\epsilon =  \frac{m}{2}
 {\vec{v}_{A(s)}}^2 
 +E_{\gamma(s)},
\end{eqnarray}
where  
$E_{\gamma(s)} = h \nu_{(s)}$, and 
$\vec{P}_{\gamma(s)} =( h \nu_{(s)}/c_{(s)}) \hat{n}$; but $c_{(s)}$ is unknown yet, as the source frame is not the absolute frame.

We first compare the momentum conservation formulas in Eqs.~(\ref{Eq.2-1}) and (\ref{Eq.2-3}). 
From the Galilean velocity transformation, we have 
\begin{equation}
\label{Eq.2-5}
\vec{v}_{A(d)} = \vec{v}_{A(s)} +  \vec{v}.
\end{equation} 
Substituting Eq.~(\ref{Eq.2-5}) into  Eq.~(\ref{Eq.2-1}), we have
\begin{equation}
\label{Eq.2-6}
0 = m \vec{v}_{A(s)} 
+\vec{P}_{\gamma (d)}.
\end{equation} 
From Eqs.~(\ref{Eq.2-6}) and (\ref{Eq.2-3}), we see that 

\begin{equation}
\label{Eq.2-7}
\vec{P}_{\gamma (s)}
 = \vec{P}_{\gamma (d)}.
\end{equation} 
This shows that the momentum of light quanta does not transform when switching reference frames. This is consistent with the formula in Eq.~(\ref{Eq.1-6}). 

Next we compare the energy conservation formulas. 
Substituting  Eq.~(\ref{Eq.2-5}) to Eq.~(\ref{Eq.2-2}), and using Eq.~(\ref{Eq.2-4}), we obtain  

\begin{equation}
\label{Eq.2-8}
 E_{\gamma(d)}
 = 
 E_{\gamma(s)}- m \vec{v}_{A(s)} \cdot \vec{v}.
\end{equation} 
Using Eq.~(\ref{Eq.2-6}), Eq.~(\ref{Eq.2-8}) can be written as:
\begin{eqnarray}
E_{\gamma(d)}
&=&
 E_{\gamma(s)}+ \vec{P}_{\gamma (d)} \cdot \vec{v}
 \nonumber
 \\
 \label{Eq.2-9}
 &=&
 E_{\gamma(s)}+ 
 E_{\gamma (d)}
 \frac{  \vec{v} \cdot \hat{n}}{c}. 
\end{eqnarray}
Thus,
\begin{equation}
\label{Eq.2-10}
E_{\gamma(d)} 
=
\frac{1}{(1- \frac{\vec{v} \cdot \hat{n}}{c})} E_{\gamma(s)}.
\end{equation} 
As $E=h \nu$,
we then have
\begin{equation}
\label{Eq.2-11}
\nu_{(d)} 
=
\frac{1}{(1- \frac{\vec{v} \cdot \hat{n}}{c})} \nu_{(s)}.
\end{equation} 
This gives the exact classical Doppler effect for the case of a moving source.

From  Eq.~(\ref{Eq.2-7}), we have 
$\nu_{(s)}/c_{(s)}= \nu_{(d)}/c $. 
Substituting this into Eq.~(\ref{Eq.2-11}), we obtain that the velocity of light quanta in the source $A^*$ frame is
\begin{equation}
\label{Eq.2-12}
c_{(s)} =
c (1- \frac{\vec{v} \cdot \hat{n}}{c})
= c- \vec{v} \cdot \hat{n}.
\end{equation} 
This result is the same as that in Eq.~(\ref{Eq.1-5}). From the view point of the wave theory,  the medium is like a wind blowing with a velocity ($-\vec{v}$) to the source $A^*$. And therefore the velocity of the light quanta relative to $A^*$ is of the value $c_{(s)}$ described in Eq.~(\ref{Eq.2-12}). 
From (\ref{Eq.2-10}) and (\ref{Eq.2-12}), the classical photon has such a property as

\begin{equation}
\label{Eq.2-13}
\frac{E_{\gamma(d)}}{E_{\gamma(s)}}
=
\frac{c}{c_{(s)}}
=
\frac{c_{(d)}}{c_{(s)}}.
\end{equation} 
Or, 
\begin{equation}
\label{Eq.2-14}
\frac{E_{\gamma(d)}} {c_{(d)}}
=
\frac{E_{\gamma(s)}} {c_{(s)}}.
\end{equation} 
Eq.~(\ref{Eq.2-14}) is just the statement that $|\vec{P}_{\gamma(d)}|
= |\vec{P}_{\gamma(s)}|$. 
Eqs.~(\ref{Eq.2-13}) and (\ref{Eq.2-14}) are only applied to classical photons, not to relativistic photons. 

Below, we discuss Fermi's derivation of the classical Doppler effect. From Eqs.~(\ref{Eq.2-3}) and (\ref{Eq.2-4}), we obtain
\begin{equation}
\label{Eq.2-15}
\epsilon
= \frac{\vec{P}^2_{\gamma(s)}}
{2m}
+ E_{\gamma(s)}.
\end{equation} 
And then, from Eqs.~(\ref{Eq.2-1}) and (\ref{Eq.2-2}), we obtain
\begin{equation}
\label{Eq.2-16}
\epsilon
= 
- \vec{v} \cdot \vec{P}_{\gamma(d)}
+
\frac{\vec{P}^2_{\gamma(d)}}
{2m}
+ E_{\gamma(d)}.
\end{equation} 
Fermi first neglected the term $\vec{P}^2_{\gamma(s)}/(2m)$ in Eq.~(\ref{Eq.2-15}) to get $\epsilon \approx h \nu_{\gamma(s)}$, 
due to $ m c^2 >> E_{\gamma(s)}$. He then substituted this result to Eq.~(\ref{Eq.2-16}). And then he next neglected the term $ \vec{P}^2_{\gamma(d)}/(2m)$ to obtained the classical Doppler effect. It is interesting to note that the two terms Fermi neglected in fact just cancellate each other, due to Eq.~(\ref{Eq.2-7}). That is why Fermi could obtain the right formula of Doppler effect by approximation.

\section{\label{sec:III} The detector is moving and the source is at rest}

We next consider the detector is moving with velocity $\vec{v}$ with respect  to the absolute frame $K$, while the source of light, $A^*$, is stationary in the $K$ frame. 
%
%

In the detector frame, the source $A^*$ is moving with velocity $\vec{v}_{A^*(d)}= - \vec{v}$.
Eqs.~(\ref{Eq.1-1})-(\ref{Eq.1-2}) in the detedtor frame are as follows:
\begin{eqnarray}
&& 
A^* \rightarrow A + \gamma
\nonumber
\\
\label{Eq.3-1}
&&
 -m \vec{v} = m \vec{v}_{A(d)}
 +\vec{P}_{\gamma (d)},
\\ 
\label{Eq.3-2}
&&
\epsilon+ 
\frac{m}{2} {\vec{v}}^2 = 
 \frac{m}{2} {\vec{v}_{A(d)} }^2 
 +E_{\gamma(d)},
\end{eqnarray}
where  
$E_{\gamma(d)} = h \nu_{(d)}$, and 
$\vec{P}_{\gamma(d)} =( h \nu_{(d)}/c_{(d)}) \hat{n}$; but $c_{(d)}$ is unknown yet, because the detector frame is not the absolute frame.

In the source frame, $A^*$ is at rest. 
Eqs.~(\ref{Eq.1-1})-(\ref{Eq.1-2}) in the source frame are
\begin{eqnarray}
\label{Eq.3-3}
&& 
A^* \rightarrow A + \gamma.
\nonumber
\\
&&
 0 = m \vec{v}_{A(s)}
 +\vec{P}_{\gamma (s)},
\\ 
\label{Eq.3-4}
&&
\epsilon =
 \frac{m}{2} {\vec{v}_{A(s)}}^2 
 +E_{\gamma(s)},
\end{eqnarray}
where  
$E_{\gamma(s)} = h \nu_{(s)}$, and 
$\vec{P}_{\gamma(s)} =( h \nu_{(s)}/c) \hat{n}$, as the source frame is the absolute frame.
From the Galilean velocity transformation, we have
\begin{equation}
\label{Eq.3-5}
\vec{v}_{A(s)} = \vec{v}_{A(d)} +  \vec{v}.
\end{equation} 
Substituting Eq.~(\ref{Eq.3-5}) into  Eq.~(\ref{Eq.3-3}), we have
\begin{equation}
\label{Eq.3-6}
0 = m \vec{v}_{A(d)} 
+m \vec{v}+
\vec{P}_{\gamma (s)}.
\end{equation} 
Comparing Eqs.~(\ref{Eq.3-6}) with  (\ref{Eq.3-1}), we see that 

\begin{equation}
\label{Eq.3-7}
\vec{P}_{\gamma (s)}
 = \vec{P}_{\gamma (d)}.
\end{equation} 
Again, this shows that the momentum of the light quanta does not transform when switching reference frames. 

Substituting  Eq.~(\ref{Eq.3-5}) to Eq.~(\ref{Eq.3-2}), and comparing with Eq.~(\ref{Eq.3-4}), we obtain 

%
%

\begin{equation}
\label{Eq.3-8}
E_{\gamma(d)}
=
E_{\gamma(s)} + m \vec{v}_{A(s)} \cdot \vec{v} 
\end{equation}
\noindent
Using  Eq.~(\ref{Eq.3-3}), we have

\begin{eqnarray}
\label{Eq.3-9}
E_{\gamma(d)}
&=&
 E_{\gamma(s)}- \vec{P}_{\gamma (s)} \cdot \vec{v}
\\
\label{Eq.3-10} 
&=&
E_{\gamma(s)}
(1- \frac{\vec{v} \cdot \hat{n}}{c}).
\end{eqnarray}
As $E = h \nu$,
we then have

\begin{equation}
\label{Eq.3-11}
\nu_{(d)} 
=
(1- \frac{\vec{v} \cdot \hat{n}}{c}
)
 \nu_{(s)}.
\end{equation} 

\noindent 
This gives the exact classical Doppler effect for the case of a moving detector.

From  Eq.~(\ref{Eq.3-7}), we then have 
$\nu_{(s)}/c_{(s)}= \nu_{(d)}/c $. 
Substituting this into Eq.~(\ref{Eq.3-11}), we then have
\begin{equation}
\label{Eq.3-12}
c_{(d)} =
c (1- \frac{\vec{v} \cdot \hat{n}}{c})
= c- \vec{v} \cdot \hat{n}.
\end{equation} 
This result is the same as that in Eq.~(\ref{Eq.1-5}). From the view point of the wave theory, the medium is like a wind blowing with a velocity ($-\vec{v}$) to the detector. And therefore the velocity of the light quanta relative to the detector is of the value $c_{(d)}$ described in Eq.~(\ref{Eq.3-12}). 
From (\ref{Eq.3-10}) and (\ref{Eq.3-12}), the classical photon has such a property as

\begin{equation}
\label{Eq.3-13}
\frac{E_{\gamma(d)}}{E_{\gamma(s)}}
=
\frac{c_{(d)}}{c}
=
\frac{c_{(d)}}{c_{(s)}}.
\end{equation} 
Again, Eq.~(\ref{Eq.3-13}) is just the statement that $|\vec{P}_{\gamma(d)}|
= |\vec{P}_{\gamma(s)}|$.

\section{\label{sec:IV} The relation of Doppler shift between two inertial reference frames}

We denote by $D_1$ and $D_2$ the two arbitrary inertial reference frames.
%
We take $D_1$ and $D_2$ as two detectors, moving with velocities $\vec{v}_1$ and $\vec{v}_2$, respectively to the frame $K$. 
Applying Eq.~(\ref{Eq.3-7}) for $D_1$ and $D_2$ detectors, we easily obtain  
\begin{equation}
\label{Eq.4-1}
\vec{P}_{\gamma(D_2)}
=
\vec{P}_{\gamma(D_1)}.
\end{equation}
Using the relation, $\vec{P} = \hbar\ \vec{k}$, Eq.~(\ref{Eq.4-1}) is just the same as that in Eq.~(\ref{Eq.1-6}). 
The same, applying Eq.~(\ref{Eq.3-13}) for $D_1$ and $D_2$ detectors, we easily obtain  
\begin{equation}
\label{Eq.4-2}
\frac
{E_{\gamma(D_2)}}
{E_{\gamma(D_1)}}
=
\frac{c_2}{c_1},
\end{equation}
where
$ c_1 = c- \vec{v}_1 \cdot \hat{n} $ is the speed of light quanta measured in the $D_1$ frame, and $ c_2 = c- \vec{v}_2 \cdot \hat{n} $, the speed of light quanta measured in the $D_2$ frame. 
Then we have
$c_2-c_1=-\vec{v} \cdot \hat{n} $,
where $\vec{v}= (\vec{v_2}-\vec{v_1})$ is the relative velocity of the two frames.
This yields

\begin{equation}
\label{Eq.4-3}
c_2 = c_1 -  \vec{v} \cdot \hat{n}.
\end{equation}
Eq.~(\ref{Eq.4-3}) is the same as that in Eq.~(\ref{Eq.1-5}). 
Also, as
$
c_2/c_1= 1+(c_2- c_1)/c_1 
= 1- \vec{v} \cdot \hat{n}/c_1
$, 
Eq.~(\ref{Eq.4-2}) can be rewritten as
\begin{equation}
\label{Eq.4-4}
E_{\gamma(D_2)}
=
(1-
\frac
{
\vec{v} \cdot \hat{n}
}
{c_1}
) E_{\gamma(D_1)},
\end{equation}

\noindent
Using the relation, $E = h \nu$, Eq.~(\ref{Eq.4-4}) is the same as that in Eq.~(\ref{Eq.1-4}). 

\section{\label{sec:V} Conclusion}

Comparing Eqs.~(\ref{Eq.1-3})-(\ref{Eq.1-6}) and Eqs.~(\ref{Eq.4-1})-(\ref{Eq.4-4}), we conclude that the classical Doppler effect deduced from photon emission process is exactly the same as those derived from the wave theory. We thus have shown that the concept of photon can also be applied to deriving the classical  Doppler theory, just the same as it was first introduced to deriving the relativistic Doppler effect.
The photon is introduced as a particle with momentum $\vec{P}$ and energy $E$ and is associated with the wave-particle duality  relations:
$E = \hbar \omega $, $\vec{P}= \hbar \vec{k}$, 
and  the relation
$\vec{P} = (E/c )\ \hat {n}$, where $c=\omega/ k$.
These relations are applied to both the relativistic and classical theory.
However, the speed $c$ is a universal constant for relativistic photons, but is frame dependent for classical photons.

With theses properties of classical and relativistic photons, and also the laws of conservation of momentum and energy, the exact classical Doppler effect and the exact relativistic Doppler effect can be derived  from the photon theory.
Interestingly, these results are the same as those derived from the wave theory.
\\
\\
\\
\\
\\

 \begin{center}
 {\bf APPENDIX}
 \end{center}
In what follows, we derive the relativistic Doppler effect. 
In general, the relativistic Doppler effect can easily be derived from the assumption that $(E_{\gamma}, \vec{P}_{\gamma}) $ forms a four-vector. 
However, we below give a derivation from the photon emission process for pedagogical purpose.
We consider that the source $A^*$ is moving at a velocity $\vec{v}$ with respect to the detector frame. We let $m_0$ be the rest mass of the atom $A$, and $m^*_0$ the rest mass of the excited atom $A^*$.
Then 
$ m^*_0 c^2 = m_0 c^2 + \epsilon$.
%
In the source frame, 
$A^*$ is at rest, the conservation of momentum and energy formulas for the photon emission are
\begin{eqnarray}
&&
A^* \rightarrow A + \gamma  
\nonumber
\\
\label{Eq.A-1}
&&
 0 =  \vec{P}_{A(s)}
 +\vec{P}_{\gamma(s)},
\\ 
\label{Eq.A-2}
&&
\epsilon + m_0 c^2=  
E_{A(s)}+
E_{\gamma(s)}.
\end{eqnarray}
While, in the detector frame, we have
\begin{eqnarray}
\label{Eq.A-3}
&&
 \vec{P}_{A*(d)} = \vec{P}_{A(d)}
 +\vec{P}_{\gamma(d)},
\\ 
\label{Eq.A-4}
&&
E_{A^*(d)} = E_{A(d)}
 +E_{\gamma(d)},
\end{eqnarray}

\noindent
As the source $A^*$ is moving at a velocity $\vec{v}$ in the detector frame,  then 
$E_{A^*(d)}= \gamma(v) (\epsilon +m_0 c^2)$,
and
$\vec{P}_{A^*(d)} = \gamma(v) (\epsilon +m_0 c^2) \vec{v}/c^2$, 
where 
$\gamma(v)= 1/ \sqrt{1-v^2/c^2}$.
We are to relate $E_{\gamma(s)}$ and $E_{\gamma(d)}$.
This relation is to be expressed in terms of quantities labelled by $A^*$. We thus should eliminate quantities labelled by $A$. 

From Eqs.~(\ref{Eq.A-1})-(\ref{Eq.A-2}), we can eliminate $E_{A(s)}$ and $\vec{P}_{A(s)}$, 
by using the formula: $E^2_{A(s)} - {\vec{P}_{A(s)} }^2 c^2  = {m_0} ^2 c^4$.
Then, we obtain
\begin{equation}
\label{Eq.A-5}
\epsilon^2+
2  m_0 c^2 \epsilon
=
2  (\epsilon + m_0 c^2) E_{\gamma(s)}.
\end{equation}
Above, we have used  $E^2_{\gamma(s)} - {\vec{P}_{\gamma(s)} }^2 c^2  = 0$.
Eq.~(\ref{Eq.A-5}) shows the relation of $E_{\gamma(s)}$ and $\epsilon$  from the view point of $A^*$. We note that when $\epsilon << m_0 c^2$, then $E_{\gamma(s)} \approx \epsilon$. This corresponds to the neglect of the recoil energy of the atom.

The same, from Eqs.~(\ref{Eq.A-3})-(\ref{Eq.A-4}), we can eliminate $E_{A(d)}$ and $\vec{P}_{A(d)}$, 
by using the formula:
$E^2_{A(d)} - {\vec{P}_{A(d)} }^2 c^2  = {m_0} ^2 c^4$.
We also need of using the formulas: 
$E^2_{A^*(d)} - {\vec{P}_{A^*(d)} }^2 c^2  = (\epsilon +m_0 c^2)^2$
and
$E^2_{\gamma(d)} - {\vec{P}_{\gamma(d)} }^2 c^2  = 0$.
Then, we obtain

\begin{equation}
\label{Eq.A-6}
\epsilon^2+
2  m_0 c^2 \epsilon
=
2 (
E_{A^*(d)} E_{\gamma(d)}
-\vec{P}_{A^*(d)} \cdot \vec{P}_{\gamma(d)} c^2
). 
\end{equation}

\noindent Substituting
$E_{A^*(d)}= \gamma(v) (\epsilon +m_0 c^2)$,
$\vec{P}_{A^*(d)} = \gamma(v) (\epsilon +m_0 c^2) \vec{v}/c^2$, 
and also
$\vec{P}_{\gamma(d)} = E_{\gamma (d)} \hat{n}/c $
to Eq.~(\ref{Eq.A-6}), we have

\begin{equation}
\label{Eq.A-7}
\epsilon^2+
2  m_0 c^2 \epsilon
=
2  (\epsilon + m_0 c^2)
\gamma(v) 
 (1 - \frac{\vec{v} \cdot \hat{n}}{c}) E_{\gamma(d)},
\end{equation}

\noindent Eq.~(\ref{Eq.A-7}) shows the relation of $E_{\gamma(d)}$, the velocity of the excited atom $\vec{v}$, and $\epsilon$, from the view point of the detector.
Comparing Eq.~(\ref{Eq.A-7}) with (\ref{Eq.A-5}), we have
\begin{equation}
\label{Eq.A-8}
E_{\gamma(d)}=
\frac{1}{ \gamma(v) (1 - \frac{\vec{v} \cdot \hat{n}}{c})} 
E_{\gamma(s)}.
\end{equation}
Using $E = h \nu$, we have

\begin{equation}
\label{Eq.A-9}
\nu_{(d)}=
\frac{1}{ \gamma(v) (1 - \frac{\vec{v} \cdot \hat{n}}{c})}
\nu_{(s)}.
\end{equation}
This gives the exact relativistic Doppler effect.

\begin{acknowledgments}
The author is indebted to Prof. Tsin-Fu Jiang and Prof. Young-Sea Huang for many help and discussions.
\end{acknowledgments}

\end{document}